\documentclass{svproc}
\usepackage{url}

\usepackage{graphicx}
\usepackage{multicol}
\usepackage{footmisc}
\usepackage{amsmath,amssymb,amsfonts}
\usepackage{caption}
\usepackage{subcaption}
\usepackage{hyperref}
\usepackage{xcolor}

\begin{document}
\title{Study of the internal structure of the Earth using neutrino oscillations at IceCube DeepCore }
%
\titlerunning{Study of the internal structure of the Earth}
\author{Sharmistha Chattopadhyay\footnote[2]{also at Institute of Physics, Sachivalaya Marg, Sainik School Post, Bhubaneswar 751005, India and Homi Bhabha National Institute, Anushakti Nagar, Mumbai, India.}  \\
	(For the IceCube Collaboration\footnote[1]{\url{http://icecube.wisc.edu}})}

\authorrunning{Sharmistha Chattopadhyay (For the IceCube Collaboration)} 
%
\tocauthor{For the IceCube Collaboration}
\institute{Dept. of Physics and Wisconsin IceCube Particle Astrophysics Center, University of Wisconsin-Madison, Madison, WI 53706, USA \\
	\email{schattopadhyay@icecube.wisc.edu}}

\maketitle              
\begin{abstract}
	
Earth’s mass and internal structure have been primarily studied through gravitational and seismic methods. Neutrinos, however, offer an independent way to explore Earth’s interior via matter effects in neutrino oscillations that depend on the electron distribution inside Earth, and hence its matter density. Our study uses atmospheric neutrinos at DeepCore, a densely instrumented sub-detector of the IceCube Neutrino Observatory, to estimate Earth’s mass and layer densities. We also assess how the upcoming IceCube Upgrade, with denser instrumentation, could improve these measurements. 

\end{abstract}
\section{Introduction}

Earth's deep internal structure remains a challenging scientific puzzle due to extreme temperatures and pressures that limit direct exploration. Indirect methods use seismic studies~\cite{Robertson:1966} and gravitational measurements~\cite{astro_almanac} to infer details about Earth's mass, moment of inertia, composition, and internal layers. Despite advances, key uncertainties persist, particularly in understanding density variations between the inner and outer core, the core's chemical composition, and unknown quantities of light elements in the mantle and core. Neutrino-based methods offer a novel and complementary approach to probing Earth’s interior. While early ideas focused on using the absorption of high-energy neutrinos, which causes significant attenuation as the neutrinos traverse Earth's diameter at energies above a few TeV, another approach likely involves neutrino oscillations. The refinement in the measurement of the oscillation parameters, especially $\theta_{13}$, has opened up new avenues for such studies. Neutrinos crossing Earth undergo Mikheyev–Smirnov–Wolfenstein (MSW)~\cite{Mikheev:1986gs} and parametric resonances~\cite{Akhmedov:1998ui}, enabling us to study Earth's internal structure. Atmospheric neutrinos, with their huge range of baselines and energy, provide a continuous source for such studies. In our work, we utilize atmospheric neutrino oscillations at IceCube Deepcore  to investigate Earth's internal structure, focusing on two goals: (1) estimating Earth’s mass, and (2) estimating the densities of its internal layers. We present here an analysis of  simulated DeepCore data to establish a robust setup, which will later be applied to real experimental data.

\section{Analysis}
\label{sec:detector}

The IceCube Neutrino Observatory consists of 5,160 Digital Optical Modules (DOMs) attached along 86 strings, instrumenting one cubic-kilometer of glacial ice at the South Pole. Each DOM contains a sensitive photomultiplier tube that detects Cherenkov light emitted by secondary charged particles when neutrinos interact in the ice. The majority of 78 strings are spaced coarsely to detect high-energy neutrinos (TeV-PeV), while the DeepCore sub-array, with 15 densely packed strings, detects lower-energy neutrinos (few GeV range) enhancing the study of atmospheric neutrino oscillations. Starting in 2025/26, the IceCube Upgrade, with a fiducial volume of about 2 Mton, will add seven more densely packed strings, further enhancing sensitivity to low-energy neutrinos down to $\sim$1 GeV.

In this work, we use a Monte-Carlo (MC) sample equivalent to 9.3 years of IceCube DeepCore data, and Convolutional Neural Network (CNN) trained for event reconstruction and particle identification (PID).  Since our analysis utilizes Earth's matter effect on neutrino oscillations, our signal is prominent in the lower energy and in the longer baselines. The MC sample is binned into 20 logarithmic energy bins from 3 to 100 GeV (where we exclude bins below 5 GeV to ensure analysis robustness), 20 linear zenith bins from -1.0 to 0.0, and 3 PID bins [0.0, 0.33, 0.39, 1.0], representing cascade, mixed, and track-like event topologies. Here PID values closer to 1 suggest a track-like event (likely $\nu_\mu$ charged-current interaction) and values closer to 0 suggest a cascade-like topology (mainly $\nu_e$ charged-current and all neutral-current interactions). The analyses have been performed by re-weighting the MC events with flux, oscillation probability, cross section, and detector response, including their systematic uncertainties.

Our analysis has two main objectives: (1) Measuring Earth's mass and (2) Measuring correlated densities of its layers.
\begin{enumerate}
    \item Mass of Earth: We scale the densities of all the layers of a 12-layered PREM (Preliminary Reference Earth Model) profile by a factor, $\alpha$, to vary Earth's total mass. Here, we aim to assess how well neutrinos alone (without gravitational data) can estimate mass of Earth. 
    \item Correlated density measurement: In the second part of our analysis, we measure Earth’s layer densities respecting gravitational constraints of mass and moment of inertia, based on a 5-layered averaged PREM profile, and is performed independently to the first part. The layers include the inner core, outer core, inner mantle, middle mantle, and outer mantle. The core layers share one scaling factor to maintain their density ratio to the same ratio as in the PREM model, while separate factors adjust the inner and middle mantle, with the outer mantle fixed due to its low-density uncertainty. Though three scaling factors are used, gravitational constraints of mass and moment of inertia of Earth introduce correlations, allowing us to parametrize the variation in terms of a single independent scaling factor, here, taken as the scaling factor of core ($\alpha_c$).
\end{enumerate}

The parameters $\alpha$ and $\alpha_c$, representing Earth’s mass and correlated density measurement, are treated as observables in their respective analyses and will be measured using real DeepCore data. As each analysis can be performed by varying their corresponding observables, they have a nested nature and follow Wilks’ theorem.

\section{Asimov Sensitivity with DeepCore}
In this section, we present the Asimov sensitivity using simulated data equivalent to 9.3 years of DeepCore data. Fig. 1a shows the sensitivity to Earth’s mass using a 12-layered PREM profile, without external constraints. Fig. 1b displays the 1$\sigma$ band for correlated density measurements, obtained by combining simulated neutrino data with external constraints. The hatched region represents the density band allowed by mass and moment of inertia constraints. The orange band shows the 1$\sigma$ region from 9.3 years of simulated IceCube DeepCore data with these constraints. The dot-dashed orange line marks the 1$\sigma$ band edges, illustrating the correlation between density variations in different layers. This plot suggests that combining neutrino oscillation data with external constraints would improve the allowed density parameter space.

\section{Asimov Sensitivity with Upgrade}
We further assessed the improvement in sensitivity by adding 3 years of simulated Upgrade data to the DeepCore projections. Fig. 1c shows the mass of Earth, where the dashed line represents 15 years of IC86 (simulated DeepCore data), and the solid line shows the result from combining 12 years of IC86 with 3 years of IC93 (simulated IceCube Upgrade data). Fig. 1d compares the Asimov 1$\sigma$ bands for 15 years of IC86 (orange band) and the combined fit of 12 years of IC86 with 3 years of IC93 (green band) for the correlated density measurement. From both of these plots, we see a significant improvement in the results when we add 3 years of Upgrade to the DeepCore projections.

\begin{figure}[ht]
    \centering
    \begin{subfigure}[b]{0.45\textwidth}
        \centering
        \includegraphics[width=5.7cm, height=5.0cm]{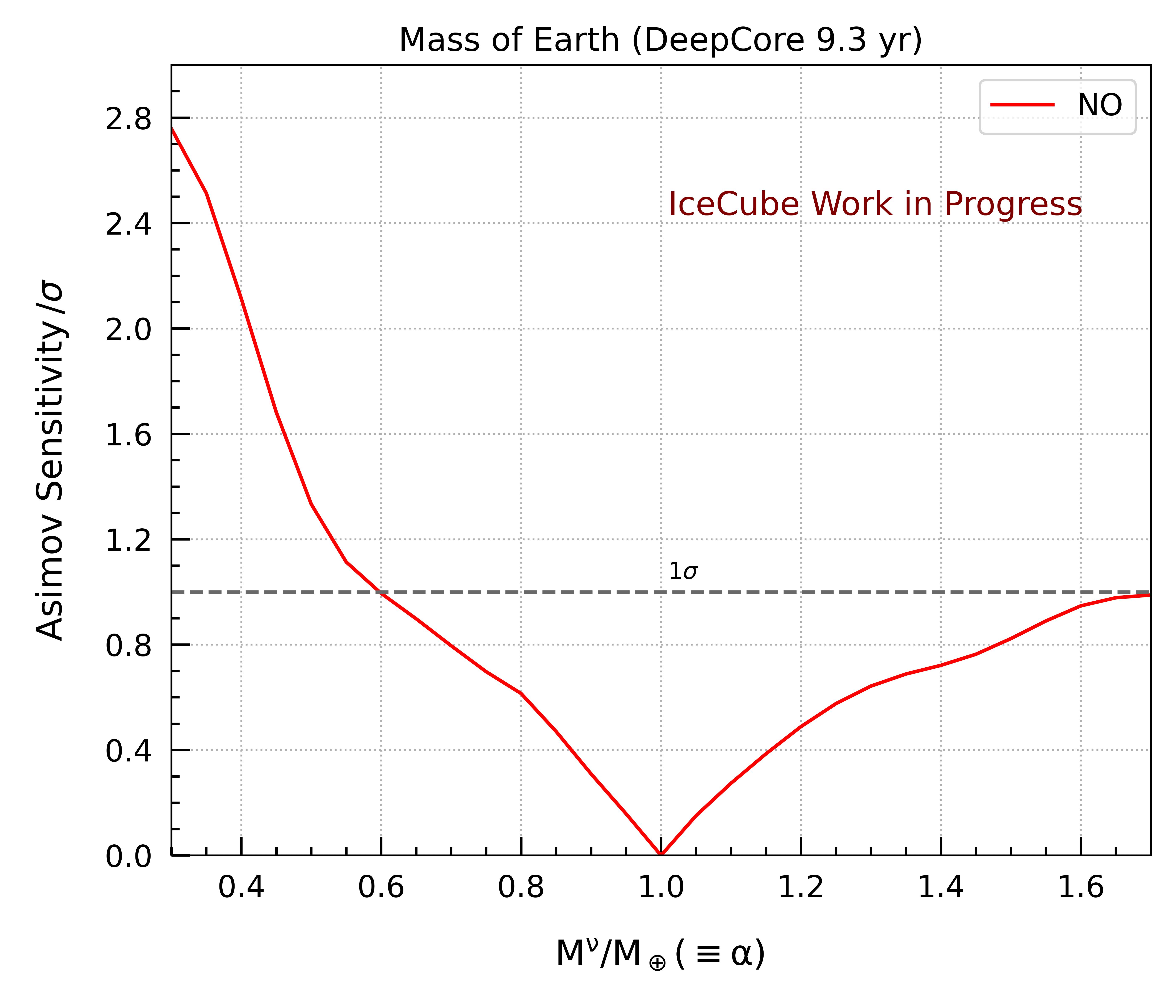} 
        \caption{}
    \end{subfigure}
    \hspace{1em} 
    \begin{subfigure}[b]{0.45\textwidth}
        \centering
        \includegraphics[width=5.7cm, height=4.9cm]{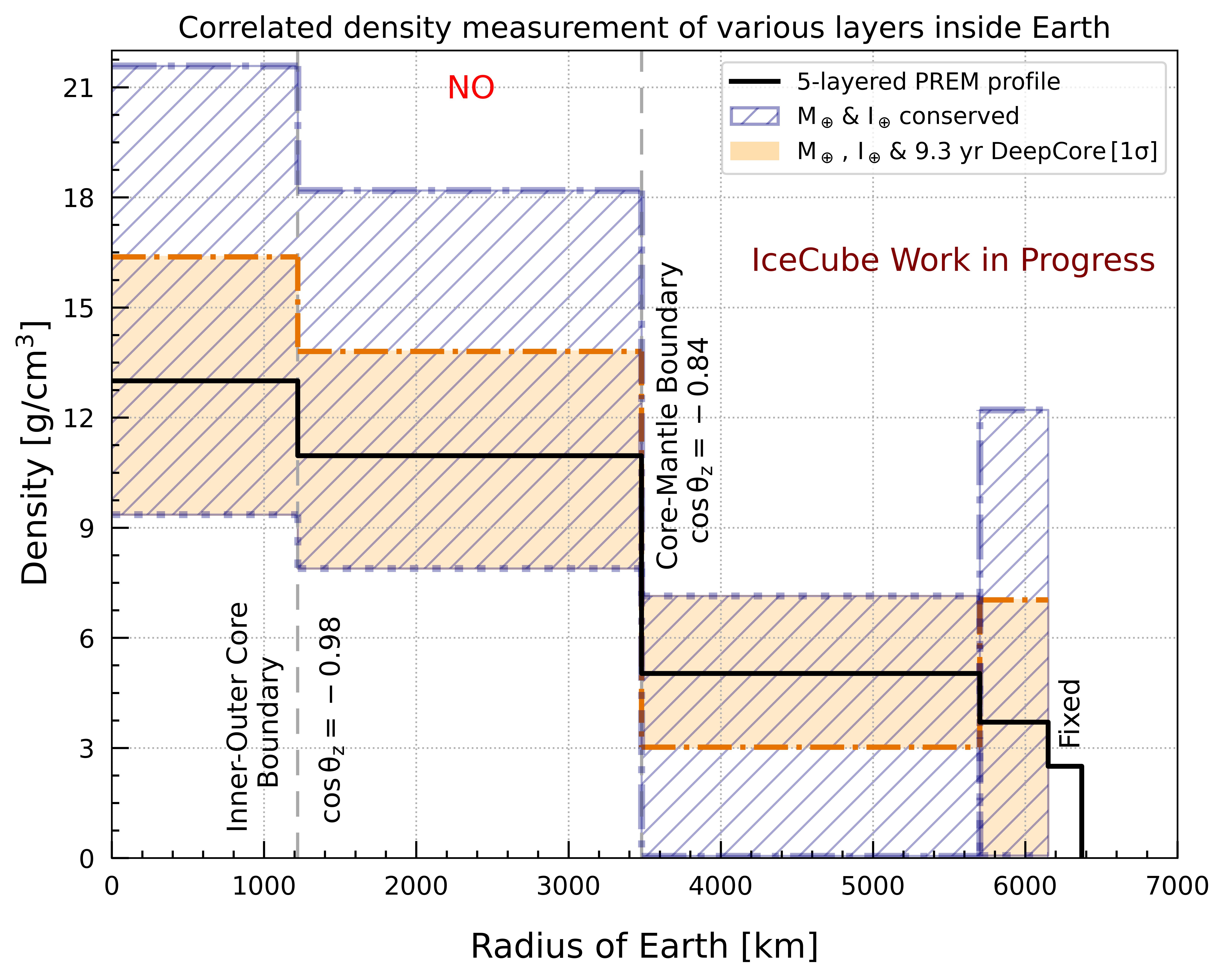} 
        \caption{}
    \end{subfigure}
    \hspace{1em}
    \begin{subfigure}[b]{0.45\textwidth}
        \centering
        \includegraphics[width=6.2cm, height=5.3cm]{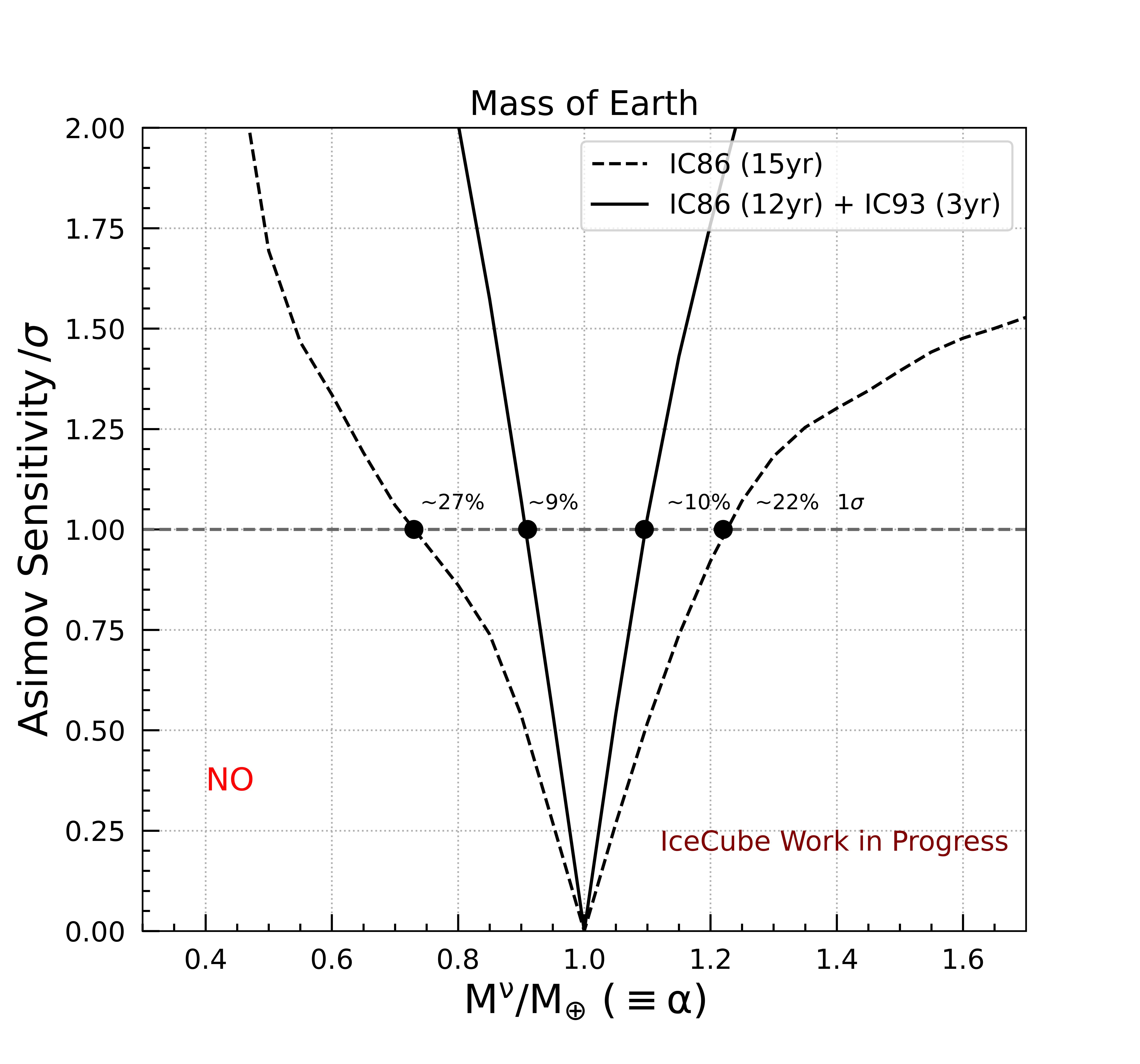} 
        \caption{}
    \end{subfigure}
    \hspace{1em}
    \begin{subfigure}[b]{0.45\textwidth}
        \centering
        \includegraphics[width=5.7cm, height=4.9cm]{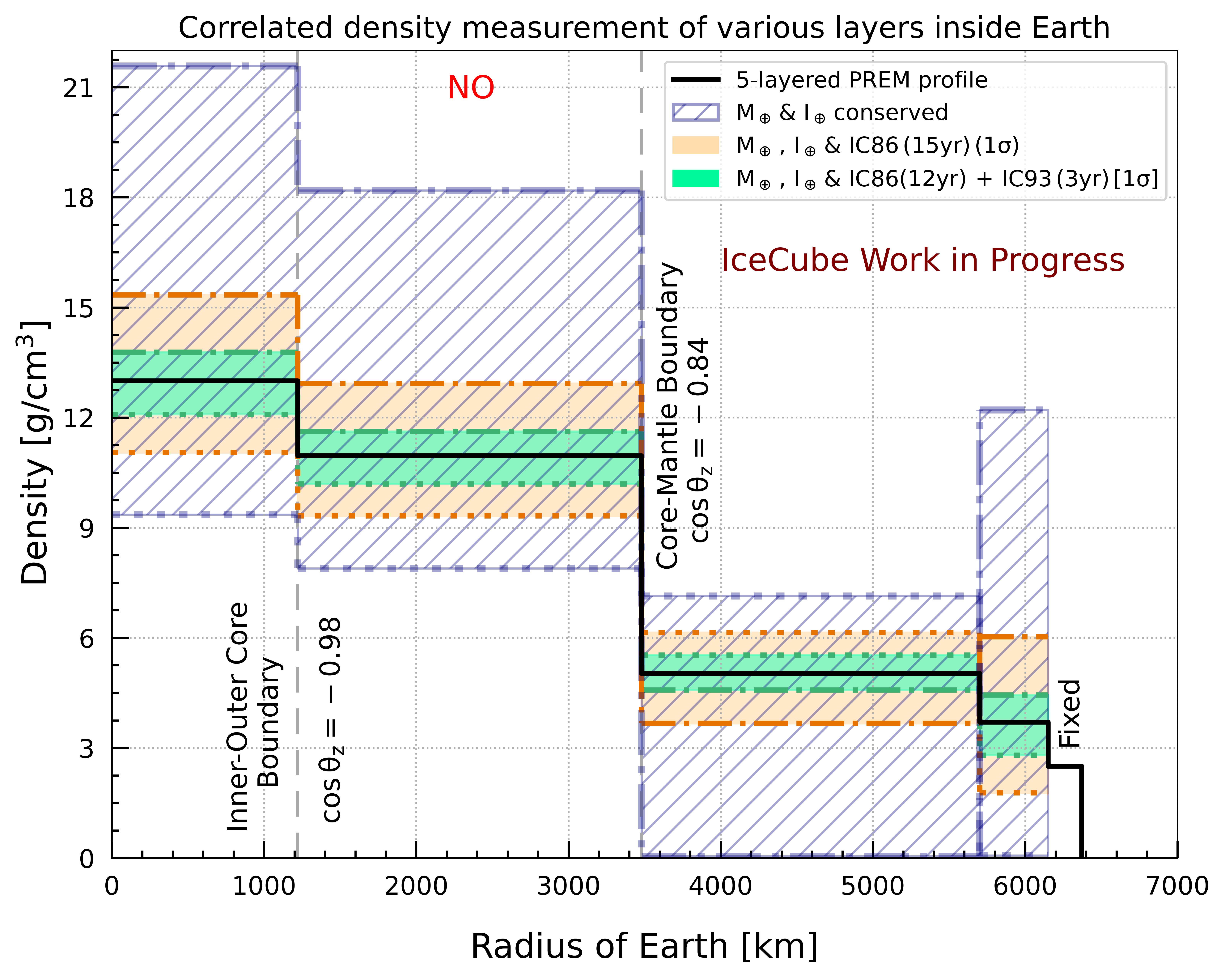} 
        \caption{}
    \end{subfigure}
    \caption{The top row shows the Asimov sensitivity results using 9.3 years of simulated DeepCore data for (a) Earth's mass and (b) correlated density measurement. The ratio $M_\nu$/$M_\oplus$ in (a) represents the scaling factor, where $M_\nu$ is the mass measured by neutrino and $M_\oplus$ is the gravitationally measured mass of Earth. The bottom row compares the Asimov sensitivities of IC86 (15 years) and IC86 (12 years) + IC93 (3 years) for (c) Earth's mass and (d) correlated density measurement.}
\end{figure}

\section{Conclusion}

In this work, we presented Asimov sensitivities for measuring the mass of Earth and the correlated densities of its internal layers using 9.3 years of simulated data at DeepCore. We have further evaluated the improvement in the sensitivity when we add 3 years of simulated Upgrade data to simulated DeepCore data. From our sensitivity results, we can see that neutrino oscillation tomography holds great potential for providing complementary insights into Earth’s internal structure. While challenges remain, advanced detectors, such as the IceCube Upgrade, will improve our measurements.

\subsubsection{\ackname} We acknowledge financial support from the Department of
Atomic Energy (DAE), Govt. of India.

%
%
%
%

\end{document}